\documentclass[12pt]{article}
\usepackage{amssymb}
\begin{document}



\def\a{\alpha}
\def\b{\beta}
\def\d{\delta}
\def\e{\epsilon}
\def\g{\gamma}
\def\h{\mathfrak{h}}
\def\k{\kappa}
\def\l{\lambda}
\def\o{\omega}
\def\p{\wp}
\def\r{\rho}
\def\t{\tau}
\def\s{\sigma}
\def\z{\zeta}
\def\x{\xi}
 \def\A{{\cal{A}}}
 \def\B{{\cal{B}}}
 \def\C{{\cal{C}}}
 \def\D{{\cal{D}}}
\def\G{\Gamma}
\def\K{{\cal{K}}}
\def\O{\Omega}
\def\R{\bar{R}}
\def\T{{\cal{T}}}
\def\L{\Lambda}
\def\f{E_{\tau,\eta}(sl_2)}
\def\E{E_{\tau,\eta}(sl_n)}
\def\Zb{\mathbb{Z}}
\def\Cb{\mathbb{C}}

\def\R{\overline{R}}

\def\beq{\begin{equation}}
\def\eeq{\end{equation}}
\def\bea{\begin{eqnarray}}
\def\eea{\end{eqnarray}}
\def\ba{\begin{array}}
\def\ea{\end{array}}
\def\no{\nonumber}
\def\le{\langle}
\def\re{\rangle}
\def\lt{\left}
\def\rt{\right}

\newtheorem{Theorem}{Theorem}
\newtheorem{Definition}{Definition}
\newtheorem{Proposition}{Proposition}
\newtheorem{Lemma}{Lemma}
\newtheorem{Corollary}{Corollary}
\newcommand{\proof}[1]{{\bf Proof. }
        #1\begin{flushright}$\Box$\end{flushright}}

\baselineskip=20pt

\newfont{\elevenmib}{cmmib10 scaled\magstep1}
\newcommand{\preprint}{
   \begin{flushleft}
   \end{flushleft}\vspace{-1.3cm}
   \begin{flushright}\normalsize
     {\tt hep-th/0511134} \\ November 2005
   \end{flushright}}
\newcommand{\Title}[1]{{\baselineskip=26pt
   \begin{center} \Large \bf #1 \\ \ \\ \end{center}}}
\newcommand{\Author}{\begin{center}
   \large \bf
Wen-Li Yang$,{}^{a,b}$
 ~ Rafael I. Nepomechie${}^c$ and~Yao-Zhong Zhang${}^b$\end{center}}
\newcommand{\Address}{\begin{center}

     ${}^a$ Institute of Modern Physics, Northwest University,
     Xian 710069, P.R. China\\
     ${}^b$ Department of Mathematics, University of Queensland, Brisbane, QLD 4072,
     Australia\\
     ${}^c$ Physics Department, P.O. Box 248046,\,University of
     Miami, Coral Gables, FL\,33124, USA
   \end{center}}
\newcommand{\Accepted}[1]{\begin{center}
   {\large \sf #1}\\ \vspace{1mm}{\small \sf Accepted for Publication}
   \end{center}}

\preprint
\thispagestyle{empty}
\bigskip\bigskip\bigskip

\Title{$Q$-operator and $T-Q$ relation from the fusion hierarchy}
\Author

\Address
\vspace{1cm}

\begin{abstract}
We propose that the Baxter $Q$-operator for the spin-1/2 XXZ
quantum spin chain is given by the $j\rightarrow \infty$ limit of
the transfer matrix with spin-$j$ (i.e., $(2j+1)$-dimensional)
auxiliary space. Applying this observation to the open chain with
general (nondiagonal) integrable boundary terms, we obtain from
the fusion hierarchy the $T$-$Q$ relation for {\it generic\/}
values (i.e. not roots of unity) of the bulk anisotropy parameter.
We use this relation to determine the Bethe Ansatz solution of the
eigenvalues of the fundamental transfer matrix.  This approach is
complementary to the one used recently to solve the same model for
the roots of unity case.

\vspace{1truecm} \noindent {\it PACS:} 03.65.Fd; 04.20.Jb;
05.30.-d; 75.10.Jm

\end{abstract}
\newpage
\section{Introduction}
\label{intro} \setcounter{equation}{0}

The Baxter $Q$-operator is a fundamental object in the theory of
exactly solvable models \cite{Bax82}.  Nevertheless, it has been
an enigma.  Indeed, while the transfer matrix has a systematic
construction in terms of solutions of the Yang-Baxter equation,
the $Q$-operator's original construction -- its brilliance
notwithstanding -- was ad hoc.  In particular, the $Q$-operator
seemed to be absent from the quantum inverse scattering method
(QISM). It was later understood \cite{Baz97, Ant97} that the
$Q$-operator could be realized by a transfer matrix whose
associated auxiliary space is infinite dimensional.  However, its
relation to the QISM remained unclear.

Motivated in part by \cite{Baz97, Ant97}, we propose here that the
$Q$-operator $\bar{\cal{Q}}(u)$ for a spin-1/2 XXZ quantum spin chain is
given by the $j\rightarrow \infty$ limit of the transfer matrix
$t^{(j)}(u)$ with spin-$j$ (i.e., $(2j+1)$-dimensional) auxiliary
space, \bea \bar{{\cal{Q}}}(u) = \lim_{j\longrightarrow
\infty}t^{(j)}(u-2j\eta) \,, \label{Qbar} \eea where $\eta$ is the
anisotropy parameter. This relation makes it clear that  the
$Q$-operator does in fact fit naturally within the QISM. Moreover,
this relation together with the fusion hierarchy for the
closed-chain transfer matrix \cite{Kar79,Kul81,Kul82,Kir87} \bea
t^{({1\over 2})}(u)\ t^{(j)}(u-2j \eta) &=&
\sinh^{N}(u+\eta)\sinh^{N}(u-\eta)\ t^{(j-{1\over
2})}(u-2(j-{1\over 2})
\eta-\eta) \no \\
&+& t^{(j+{1\over 2})}(u-2(j+{1\over 2}) \eta+\eta),\quad
j=\frac{1}{2},1,\frac{3}{2},\ldots \eea immediately leads to the
Baxter $T-Q$ relation \bea t^{({1\over 2})}(u)\ {\bar
{\cal{Q}}}(u) = \sinh^{N}(u+\eta)\sinh^{N}(u-\eta)\ {\bar
{\cal{Q}}}(u-\eta) + {\bar {\cal{Q}}}(u+\eta) \label{closedTQ} \,,
\eea from which it is possible to derive the well-known expression
for the eigenvalues of the fundamental transfer matrix
$t^{({1\over 2})}(u)$ and the associated Bethe Ansatz equations.
However, we emphasize that the above argument is formal: we assume
without proof that the limit in (\ref{Qbar}) exists, and we do not
evaluate the right-hand-side explicitly.

It is interesting to apply this observation to the open spin-1/2
XXZ quantum spin chain with general integrable boundary terms
\cite{Veg93,Gho94}.  Indeed, this model remains unsolved, although
the special case of diagonal boundary terms was solved long ago
\cite{Gau71,Alc87,Skl88}.  Significant progress has been made
recently for the case of nondiagonal boundary terms where the
boundary parameters obey some constraints.  One approach
\cite{Cao03} (see also \cite{Yan04}) is based on the generalized
algebraic Bethe Ansatz \cite{Bax73,Fad79}. \footnote{It is not yet
clear whether this approach can give all the eigenvalues of the
transfer matrix.  Indeed, in order to obtain all the levels, two
sets of Bethe Ansatz equations are required \cite{Nep03}, and
therefore, presumably two pseudovacua.  However, it is not yet
clear how to construct the second pseudovacuum.} A second
approach, which was developed in \cite{Nep04}, exploits functional
relations obeyed by the transfer matrix at roots of unity to
obtain the eigenvalues of the transfer matrix. These functional
relations are a consequence of the truncation of the fusion
hierarchy of the transfer matrix at roots of unity \cite{Baz96,
Kun98}.

In this paper, we develop a third approach, which is complementary to
the second one \cite{Nep04}.  Indeed, as in \cite{Nep04}, we make use
of the fusion hierarchy for the open XXZ chain \cite{Mez92,Zho96}.
However, here we consider instead {\it generic} values (i.e. not roots
of unity) of the bulk anisotropy parameter, for which the fusion
hierarchy does {\it not} truncate.  We instead use the relation
(\ref{Qbar}) to obtain the $T$-$Q$ relation.  We then use the latter
relation, together with some additional properties of the transfer
matrix, to determine the eigenvalues of the transfer matrix and the
associated Bethe Ansatz equations.  The expressions for the
eigenvalues are generalizations of those found in \cite{Nep04},
and this new derivation explains their validity for generic anisotropy
values \cite{Nep03}.  \footnote{Additional solutions have recently
been found at roots of unity which are {\it not} valid for generic
anisotropy values \cite{Mur05}.}

The paper is organized as follows.  In Section 2, we introduce our
notation and some basic ingredients.  In Section 3, we derive the
$T$-$Q$ relation from (\ref{Qbar}) and the fusion hierarchy of the
open XXZ chain.  Through that relation, we determine the eigenvalues of
the transfer matrix and the associated Bethe Ansatz equations in
Section 4.  Finally, we summarize our conclusions and mention some
interesting open problems in Section 5.


\section{ Transfer matrix}
\label{XXZ} \setcounter{equation}{0}

Throughout, let us fix a generic complex number $\eta$, and let
$\s^x,\,\s^y,\,\s^z$ be the  usual Pauli matrices. The well-known
six-vertex model R-matrix $R(u)\in {\rm End}(\Cb^2\otimes \Cb^2)$
is given by \bea
R(u)=\lt(\begin{array}{llll}\sinh(u+\eta)&&&\\&\sinh(u)&\sinh(\eta)&\\
&\sinh(\eta)&\sinh(u)&\\&&&\sinh(u+\eta)\end{array}\rt).
\label{r-matrix}\eea Here $u$ is the spectral parameter and $\eta$
is the so-called bulk anisotropy parameter. The R-matrix satisfies
the quantum Yang-Baxter equation
and the properties, \bea &&\hspace{-1.5cm}\mbox{ Unitarity
relation}:\,R_{1,2}(u)R_{2,1}(-u)=-\xi(u)\,{\rm id},
\quad \xi(u)=\sinh(u+\eta)\sinh(u-\eta),\label{Unitarity}\\
&&\hspace{-1.5cm}\mbox{ Crossing
relation}:\,R_{1,2}(u)=V_1R_{1,2}^{t_2}(-u-\eta)V_1,\quad
V=-i\s^y,
\label{crosing-unitarity}\\
&&\hspace{-1.5cm}\mbox{ Periodicity
property}:\,R_{1,2}(u+i\pi)=-\s_1^z\,R_{1,2}(u)\s^z_1.\label{quasi}
\eea
Here $R_{2,1}(u)=P_{12}R_{1,2}(u)P_{12}$ with $P_{12}$ being
the usual permutation operator and $t_i$ denotes transposition
in the $i$-th space. Here and below we adopt the standard
notations: for any matrix $A\in {\rm End}(\Cb^2)$, $A_j$ is an
embedding operator in the tensor space $\Cb^2\otimes
\Cb^2\otimes\cdots$, which acts as $A$ on the $j$-th space and as
identity on the other factor spaces; $R_{i,j}(u)$ is an embedding
operator of R-matrix in the tensor space, which acts as identity
on the factor spaces except for the $i$-th and $j$-th ones.

The transfer matrix $t(u)$ of the open XXZ chain with general
integrable boundary terms is given by \cite{Skl88} \bea
t(u)=tr_0\lt(K_0^+(u)T_0(u)K^-_0(u)\hat{T}_0(u)\rt),\label{trans}\eea
where $T_0(u)$ and $\hat{T}_0(u)$ are the monodromy matrices \bea
T_0(u)=R_{0,N}(u)\ldots R_{0,1}(u),\quad
\hat{T}_0(u)=R_{1,0}(u)\ldots R_{N,0}(u),\label{Mon}\eea and
$tr_0$ denotes trace over the ``auxiliary space" $0$. We consider
the most general solutions $K^{\mp}(u)$ \cite{Veg93,Gho94} to the
reflection equation and its dual \cite{Skl88,Che84}.  The matrix
elements are given respectively by \bea
K^-_{11}(u)&=&2\lt(\sinh(\a_-)\cosh(\b_{-})\cosh(u)
+\cosh(\a_-)\sinh(\b_-)\sinh(u)\rt),\no\\
K^-_{22}(u)&=&2\lt(\sinh(\a_-)\cosh(\b_{-})\cosh(u)
-\cosh(\a_-)\sinh(\b_-)\sinh(u)\rt),\no\\
K^-_{12}(u)&=&e^{\theta_-}\sinh(2u),\quad
K^-_{21}(u)=e^{-\theta_-}\sinh(2u),\label{K-matrix}\eea and
$K^+(u)=\lt.K^-(-u-\eta)\rt|_{(\a_-,\b_-,\theta_-)\rightarrow
(-\a_+,-\b_+,\theta_+)}$.
Here $\a_{\mp},\,\b_{\mp},\,\theta_{\mp}$ are the boundary
parameters which are associated with boundary interaction terms.
The K-matrices have the periodicity property: $
K^{\mp}(u+i\pi)=-\s^z\,K^{\mp}(u)\,\s^z$.  Sklyanin has shown that
the transfer matrices with different spectral parameters commute
with each other: $[t(u)\,, t(v)]=0$. This ensures the
integrability of the open XXZ chain. Furthermore, one can show
that the transfer matrix also has \bea
&&\hspace{-1.5cm}\mbox{$i\pi$-periodicity}:\,t(u+i\pi)=t(u),\label{Tran-Per}\\
&&\hspace{-1.5cm}\mbox{Crossing symmetry}:\,t(-u-\eta)=t(u),\label{Tran-Cr}\\
&&\hspace{-1.5cm}\mbox{Initial
condition}:\,t(0)=-2^3\hspace{-0.1cm}\sinh^{2N}(\eta)\cosh(\eta)
\sinh(\hspace{-0.06cm}\a_-\hspace{-0.06cm})
\cosh(\hspace{-0.06cm}\b_-\hspace{-0.06cm})
\sinh(\hspace{-0.06cm}\a_+\hspace{-0.06cm})
\cosh(\hspace{-0.06cm}\b_+\hspace{-0.06cm}){\rm id},\label{Tran-In}\\
&&\hspace{-1.5cm}\mbox{Asymptotic
 behavior}:\,t(u)\hspace{-0.1cm}\sim \hspace{-0.1cm}
-\frac{\cosh(\theta_--\theta_+)e^{\pm[(2N+4)u+(N+2)\eta]}}
{2^{2N+1}}{\rm id}\hspace{-0.1cm}+\hspace{-0.1cm}\ldots,\,{\rm
for}\,u\rightarrow\hspace{-0.1cm} \pm\infty. \label{Tran-Asy}\eea


\section{ $T$-$Q$ relation}
\label{T-QR} \setcounter{equation}{0}

We shall use the fusion procedure, which was first developed for
R-matrices \cite{Kar79,Kul81} and then later generalized  for
K-matrices \cite{Mez92,Zho96},  to obtain the Baxter $T$-$Q$
relation. The fused spin-$(j,\frac{1}{2})$ R-matrix
$(j=\frac{1}{2},1,\frac{3}{2},\ldots)$ is given by \cite{Kul81,Kul82}
\bea R_{\langle1\ldots 2j\rangle,2j+1}(u)=P^{(+)}_{1\ldots2j}
R_{1,2j+1}(u)R_{2,2j+1}(u+\eta)\cdots R_{2j,2j+1}(u+(2j-1)\eta)
P^{(+)}_{1\ldots2j},\eea where $P^{(+)}_{1\ldots2j}$ is the
completely symmetric projector.
Following \cite{Mez92,Zho96}, the fused spin-$j$ K-matrix
$K^-_{\langle1\ldots 2j\rangle}(u)$ is given by  \bea
K^-_{\langle1\ldots
2j\rangle}(u)&=&P^{(+)}_{1\ldots2j}\,\lt\{K^-_{2j}(u)R_{2j,2j-1}(2u+\eta)
K^-_{2j-1}(u+\eta)\rt.\no\\
&&\times
R_{2j,2j-2}(2u+2\eta)R_{2j-1,2j-2}(2u+3\eta)K^-_{2j-2}(u+2\eta)
\no\\
&&\times \cdots \lt.R_{2,1}(2u+(4j-3)\eta)
K^-_1(u+(2j-1)\eta)\rt\}\,P^{(+)}_{1\ldots2j}. \eea The fused
spin-$j$ K-matrix $K^+_{\langle1\ldots 2j\rangle}(u)$ is given by
\bea K^+_{\langle1\ldots 2j\rangle}(u)
=F(u|2j)\,\lt.K^-_{\langle1\ldots
2j\rangle}(-u-2j\eta)\rt|_{(\a_-,\b_-,\theta_-)\rightarrow
(-\a_+,-\b_+,\theta_+)}, \eea where the scalar functions $F(u|q)$
are given by $
F(u|q)={1/{(\prod_{l=1}^{q-1}\prod_{k=1}^{l}\xi(2u+l\eta+k\eta))}}$,
for $q=1,2,\ldots$. The fused transfer matrix $t^{(j)}(u)$
constructed with a spin-$j$ auxiliary space is given by \bea
t^{(j)}(u)=tr_{1\ldots 2j}\lt(K^+_{\langle1\ldots
2j\rangle}(u)T_{\langle1\ldots 2j\rangle}(u)K^-_{\langle1\ldots
2j\rangle}(u)\hat{T}_{\langle1\ldots 2j\rangle}(u+(2j-1)\eta)
\rt),\eea where \bea T_{\langle1\ldots 2j\rangle}(u)&=&R_{\langle
1\ldots 2j\rangle,N}(u)\ldots R_{\langle 1\ldots
2j\rangle,1}(u),\no\\
\hat{T}_{\langle1\ldots 2j\rangle}(u+(2j-1)\eta)&=&R_{\langle
1\ldots 2j\rangle,1}(u)\ldots R_{\langle 1\ldots 2j\rangle,N}(u).
\eea The transfer matrix (\ref{trans}) corresponds to the
fundamental case $j=\frac{1}{2}$, i.e., $t(u)=t^{(\frac{1}{2})}(u)
$. The fused transfer matrices constitute commutative families,
namely, \bea  \lt[t^{(j)}(u),\,t^{(k)}(v)\rt]=0. \label{Com-1-1}
\eea They satisfy a so-called fusion hierarchy
\cite{Mez92,Zho96}\bea
t^{(j)}(u-(2j-1)\eta)=t^{(j-\frac{1}{2})}(u-(2j-1)\eta)t(u)
-\frac{\Delta(u-\eta)}{\xi(2u)} \,
t^{(j-1)}(u-(2j-1)\eta),\label{fusion-1}\eea where $\xi(u)$ is
given by (\ref{Unitarity});  and the coefficient function
$\frac{\Delta(u-\eta)}{\xi(2u)}$, which we now denote by $\d(u)$,
is given by \footnote{In \cite{Nep04,Mur05}, the function
$\frac{\Delta(u-\eta)}{\xi(2u)}$ is denoted instead by
$\d(u-\eta)$.} \bea \d(u)=\frac{\Delta(u-\eta)}{\xi(2u)} &=&
2^4\sinh^{2N}(u-\eta)\sinh^{2N}(u+\eta)\frac{\sinh(2u-2\eta)
\sinh(2u+2\eta)}{\sinh(2u-\eta)\sinh(2u+\eta)}\no\\
&&\quad\times\sinh(u+\a_-)\sinh(u-\a_-)
\cosh(u+\b_-)\cosh(u-\b_-)\no\\
&&\quad\times \sinh(u+\a_+)\sinh(u-\a_+)
\cosh(u+\b_+)\cosh(u-\b_+). \label{Coe-function}
\eea
In the above
hierarchy, $t^{(0)}(u)={\rm id}$. For generic $\eta$, the fusion
hierarchy does not truncate (c.f. the roots of unity case \cite{Nep04}).
Hence $\{t^{(j)}(u)\}$  constitute an infinite hierarchy, namely,
$j$ taking values $\frac{1}{2},1,\frac{3}{2},\ldots$.

The commutativity (\ref{Com-1-1}) of the fused transfer matrices
$\{t^{(j)}(u)\}$ and the fusion relation (\ref{fusion-1}) imply
that the corresponding eigenvalue of the transfer matrix
$t^{(j)}(u)$, denoted by $\L^{(j)}(u)$, satisfies the following
hierarchy \bea
\L^{(j)}(u+\eta-2j\eta)=\L^{(j-\frac{1}{2})}(u-2(j-\frac{1}{2})\eta)\, \L(u)
-\d(u) \, \L^{(j-1)}(u-\eta-2(j-1)\eta). \label{fusion-2}\eea Here
we have used the convention $\L(u)=\L^{(\frac{1}{2})}(u)$ and
$\L^{(0)}(u)=1$.
Dividing both sides of (\ref{fusion-2}) by
$\L^{(j-\frac{1}{2})}(u-2(j-\frac{1}{2})\eta)$, we
have \bea \L(u)=\frac{\L^{(j)}(u+\eta-2j\eta)}
{\L^{(j-\frac{1}{2})}(u-2(j-\frac{1}{2})\eta)}+\d(u)\,
\frac{\L^{(j-1)}(u-\eta-2(j-1)\eta)}
{\L^{(j-\frac{1}{2})}(u-2(j-\frac{1}{2})\eta)}.\label{fusion-3}\eea

We now consider the limit $j \rightarrow \infty$. We make the
fundamental assumption (\ref{Qbar}) (in particular,  that the limit
exists), which implies for the corresponding eigenvalues
\bea
\bar{Q}(u)=\lim_{j\longrightarrow
+\infty}\L^{(j)}(u-2j\eta) \,.
\eea
It follows from  (\ref{fusion-3}) that
\bea
\L(u)=\frac{\bar{Q}(u+\eta)}{\bar{Q}(u)}+
\d(u)\,\frac{\bar{Q}(u-\eta)}{\bar{Q}(u)}.\label{fusion-4}\eea
Assuming the function $\bar{Q}(u)$ has the decomposition
$\bar{Q}(u)=f(u)Q(u)$ with \bea
Q(u)=\prod_{j=1}^{M}\sinh(u-u_j)\sinh(u+u_j+\eta),
\label{fusion-5}
\eea
$M$ being an integer such that $M\geq 0$,
Eq. (\ref{fusion-4}) becomes
\bea \L(u)=
H_1(u)\,\frac{Q(u+\eta)}{Q(u)}+H_2(u)\,
\frac{Q(u-\eta)}{Q(u)}.\label{T-Q}\eea Here
$H_1(u)=\frac{f(u+\eta)}{f(u)}$ and
$H_2(u)=\d(u)\,\frac{f(u-\eta)}{f(u)}$. It is easy to see that the
functions $\{H_i(u)|i=1,2\}$ satisfy the relation
\bea
H_1(u-\eta)H_2(u)=\d(u),
\label{T-Q-1}
\eea where the function
$\d(u)$ is given by (\ref{Coe-function}).

In summary, the eigenvalue $\L(u)$ of the fundamental transfer matrix
$t(u)$ (\ref{trans}) has the decomposition form (\ref{T-Q}), where the
coefficient functions $\{H_i(u)\}$ satisfy the constraint
(\ref{T-Q-1}).  In the next section, we use the analytic property of
the eigenvalue $\L(u)$ and the other properties derived from the
transfer matrix to determine the functions $\{H_i(u)\}$ and therefore
the eigenvalue $\L(u)$.


\section{ Eigenvalues and Bethe Ansatz equations}
\label{BAE} \setcounter{equation}{0}

It follows from (\ref{Tran-Per})-(\ref{Tran-Asy}) that the
eigenvalue $\L(u)$, as a function of $u$, has the following
properties, \bea
&&\hspace{-1.5cm}\mbox{Periodicity}:\,\quad\L(u+i\pi)=\L(u),\label{Eign-Per}\\
&&\hspace{-1.5cm}\mbox{Crossing
symmetry}:\,\quad\L(-u-\eta)=\L(u),\label{Eigen-Cro}\\
&&\hspace{-1.5cm}\mbox{Initial
condition}:\,\L(0)=-2^3\sinh^{2N}(\eta)\cosh(\eta)
\sinh(\a_-)\cosh(\b_-)\sinh(\a_+)\cosh(\b_+),\label{Eigen-In}\\
&&\hspace{-1.5cm}\mbox{Asymptotic behavior}:\,
\L(u)\sim-\frac{\cosh(\theta_--\theta_+)e^{\pm[(2N+4)u+(N+2)\eta]}}
{2^{2N+1}}+\ldots,\quad {\rm for}\, u\rightarrow
\pm\infty.\label{Eigen-Asy} \eea The commutativity of the transfer
matrix $t(u)$ and the analyticity of the R-matrix and K-matrices
imply that $\L(u)$ further obeys  the property \bea
\hspace{-5.2cm}\mbox{Analyticity}:\,\quad \L(u) \mbox{ is an
analytic function of $u$ at finite $u$}.\label{Eigen-Anal}\eea
Moreover the semiclassical property of the R-matrix,
$\lt.R(u)\rt|_{\eta=0}=\sinh(u)\,{\rm id}$, leads to the following
property of $\L(u)$, \bea
\lt.\L(u)\rt|_{\eta=0}&=&2^3\sinh^{2N}(u)\lt\{-\sinh(\a_-)
\cosh(\b_-)\sinh(\a_+)\cosh(\b_+)\cosh^2(u) \rt.\no \\&&\quad
+\cosh(\a_-) \sinh(\b_-)\cosh(\a_+)\sinh(\b_+)\sinh^2(u)\no\\
&&\quad \lt.-\cosh(\theta_--\theta_+)
\sinh^2(u)\cosh^2(u)\rt\}.\label{Semi}\eea

The $T$-$Q$ relations (\ref{T-Q}) and (\ref{T-Q-1}), together with
the above properties (\ref{Eign-Per})-(\ref{Eigen-Anal}),  can be
used to determine the eigenvalues of the transfer matrix. For
$\{\e_i=\pm 1\,|\,i=0,1,2,3\}$, let us introduce \bea
H^{(\pm)}_1(u|\e_1,\e_2,\e_3)&=&-2^2\e_2\sinh^{2N}(u)\frac{\sinh(2u)}
{\sinh(2u+\eta)}\sinh(u\pm\a_-+\eta)
\cosh(u\pm\e_1\b_-+\eta)\no\\
&&\quad\quad\times\sinh(u\pm\e_2\a_++\eta)\cosh(u\pm\e_3\b_++\eta),
\label{H-1}\\
H^{(\pm)}_2(u|\e_1,\e_2,\e_3)&=&-2^2\e_2\sinh^{2N}(u+\eta)
\frac{\sinh(2u+2\eta)} {\sinh(2u+\eta)}\sinh(u\mp\a_-)
\cosh(u\mp\e_1\b_-)\no\\
&&\quad\quad\times\sinh(u\mp\e_2\a_+)\cosh(u\mp\e_3\b_+).
\label{H-2}\eea One may readily check that the functions
$H^{(\pm)}_1(u|\e_1,\e_2,\e_3)$ and
$H^{(\pm)}_2(u|\e_1,\e_2,\e_3)$ indeed satisfy (\ref{T-Q-1}),
namely,  \bea
H^{(\pm)}_1(u-\eta|\e_1,\e_2,\e_3)H^{(\pm)}_2(u|\e_1,\e_2,\e_3)=
\d(u). \eea The general solution to (\ref{T-Q-1}) can be written
as follows: \bea
H_1(u)=H^{(\pm)}_1(u|\e_1,\e_2,\e_3)\,g_1(u),\quad
H_2(u)=H^{(\pm)}_2(u|\e_1,\e_2,\e_3)\,g_2(u),\eea where
$\{g_i(u)\}$ satisfy the following relations, \bea
g_1(u-\eta)g_2(u)=1,\quad g_1(u+i\pi)=g_1(u),\quad
g_2(u+i\pi)=g_2(u).\label{Eq}\eea The solutions to (\ref{Eq}) have
the following form, \bea
g_1(u)=a\frac{\prod_{j=1}^{N_2}\sinh(u-u^+_j)}
{\prod_{j=1}^{N_1}\sinh(u-u^-_j)},\quad
g_2(u)=\frac{1}{a}\frac{\prod_{j=1}^{N_1}\sinh(u-u^-_j-\eta)}
{\prod_{j=1}^{N_2}\sinh(u-u^+_j-\eta)},\eea where $N_1$ and $N_2$
are integers such that $N_1,\,N_2\geq 0$, and $a$ is an non-zero
constant. In the above equation, we assume that $u^-_j\neq
\mp\a_--\eta,\mp\e_1\b_--\eta-i\frac{\pi}{2},\mp\e_2\a_+-\eta,
\mp\e_3\b_+-\eta-i\frac{\pi}{2}$ and $u^+_j\neq
\e_0\a_-,\e'_1\b_--i\frac{\pi}{2},\e_2'\a_+,
\e'_3\b_+-i\frac{\pi}{2}$, otherwise the corresponding factors in
$g_i(u)$  make transitions among
$\{H^{(\pm)}_1(u|\e_1,\e_2,\e_3)\}$
($\{H^{(\pm)}_2(u|\e_1,\e_2,\e_3)\}$ respectively). Then the
analyticity of $\L(u)$ (\ref{Eigen-Anal}) requires that $g_{1}(u)$
and $g_{2}(u)$ have common poles; i.e., $N_1=N_2$, and
$u^-_j=u^+_{j'}+\eta$. This means \bea
g_1(u)=a\prod_{j=1}^{N_1}\frac{\sinh(u-u^-_j+\eta)}
{\sinh(u-u^-_j)},\quad
g_2(u)=\frac{1}{a}\prod_{j=1}^{N_1}\frac{\sinh(u-u^-_j-\eta)}
{\sinh(u-u^-_j)}.\eea Since
$H^{(\pm)}_2(u|\e_1,\e_2,\e_3)=H^{(\pm)}_1(-u-\eta|\e_1,\e_2,\e_3)$,
the crossing symmetry of $\L(u)$ (\ref{Eigen-Cro}) implies that
$H_2(u)=H_1(-u-\eta)$. Hence, $N_1$ is even and \bea
g_1(u)&=&a\prod_{j=1}^{\frac{N_1}{2}}\frac{\sinh(u-u^-_j+\eta)\sinh(u+u_j^-+2\eta)}
{\sinh(u-u^-_j)\sinh(u+u^-_j+\eta)},\\
g_2(u)&=&a\prod_{j=1}^{\frac{N_1}{2}}\frac{\sinh(u-u^-_j-\eta)\sinh(u+u_j^-)}
{\sinh(u-u^-_j)\sinh(u+u^-_j+\eta)},\quad a=\pm 1.\eea This is
equivalent to having additional Bethe roots; and the corresponding
factors, except $a$, can be absorbed into those of $Q(u)$
(\ref{fusion-5}). Therefore the eigenvalues of the transfer matrix
can be uniquely expressed in the following form: \bea \L(u)=a
H^{(\pm)}_1(u|\e_1,\e_2,\e_3)\,\frac{Q(u+\eta)}{Q(u)}
+aH^{(\pm)}_2(u|\e_1,\e_2,\e_3)\,\frac{Q(u-\eta)}{Q(u)},\quad
a=\pm 1.\label{T-Q-3}\eea The initial condition (\ref{Eigen-In})
implies $a=+1$. The asymptotic behavior (\ref{Eigen-Asy}),
together with (\ref{T-Q-3}), requires that the boundary parameters
should obey a constraint among the boundary parameters. The
resulting constraint and the semiclassical property of the
eigenvalues (\ref{Semi}) finally give rise to a further constraint
among the discrete parameters $\{\e_i\}$. Indeed, if the boundary
parameters satisfy any of the following constraints: \bea
\a_-+\e_1\b_-+\e_2\a_++\e_3\b_+=\e_0(\theta_--\theta_+)+\eta
k+\frac{1-\e_2}{2}i\pi\,\, {\rm mod}\, (2i\pi),\,\,
\e_1\e_2\e_3=+1, \label{Restr-1}\eea where $k$ is an integer such
that \bea |k|\leq N-1,\,\,\mbox{and}\,\, N-1+k \,\,{\rm is
~even},\label{Restr-2}\eea then the eigenvalues of the
corresponding transfer matrix are \bea
\hspace{-0.2cm}\L^{(\pm)}(u)\hspace{-0.1cm}=\hspace{-0.1cm}
H^{(\pm)}_1(u|\e_1,\e_2,\e_3)\frac{Q^{(\pm)}(u+\eta)}
{Q^{(\pm)}(u)}\hspace{-0.1cm}+\hspace{-0.1cm}
H^{(\pm)}_2(u|\e_1,\e_2,\e_3)
\frac{Q^{(\pm)}(u-\eta)}{Q^{(\pm)}(u)}. \label{result} \eea Here
\bea Q^{(\pm)}(u)=\prod_{j=1}^{M^{(\pm)}}
\sinh(u-v^{(\pm)}_j)\sinh(u+v^{(\pm)}_j+\eta),\,\,
M^{(\pm)}=\frac{1}{2}\lt(N-1\mp k\rt), \label{Eign-fuc} \eea and
the parameters $\{v^{(\pm)}_j\}$ satisfy the associated Bethe
Ansatz equations respectively, \bea
&&\frac{H^{(\pm)}_{2}(v^{(\pm)}_j|\e_1,\e_2,\e_3)}
{H^{(\pm)}_{2}(-v^{(\pm)}_j-\eta|\e_1,\e_2,\e_3)}
=-\frac{Q^{(\pm)}(v^{(\pm)}_j+\eta)}{Q^{(\pm)}(v^{(\pm)}_j-\eta)},
\quad j=1,\ldots,M^{(\pm)}.\label{BAE-2} \eea One can verify that,
for generic values of $\eta$,  both $\L^{(\pm)}(u)$ have the same
desirable asymptotic behavior (\ref{Eigen-Asy}) and semiclassical
property (\ref{Semi}) provided the constraint (\ref{Restr-1}) is
satisfied. We note that Refs. \cite{Nep04,Nep03} treat
explicitly only the case $\epsilon_1 = \epsilon_2 = \epsilon_3 = 1$.
One can check numerically along the lines of \cite{Nep03}
that for a given set of
bulk and boundary parameters satisfying (\ref{Restr-1}), the
eigenvalues $\L^{(-)}(u)$ and $\L^{(+)}(u)$ {\it together} give
the complete set of eigenvalues of the transfer matrix $t(u)$.


\section{Conclusions}
\label{Con} \setcounter{equation}{0}

We have argued that the Baxter $Q$-operator for the spin-1/2 XXZ
chain is given by the $j \rightarrow \infty$ limit of the transfer
matrix with spin-$j$ auxiliary space (\ref{Qbar}).  Indeed, this
relation together with the fusion hierarchy lead to the Baxter
$T$-$Q$ relation for both the closed (\ref{closedTQ}) and open
(\ref{T-Q}), (\ref{T-Q-1}) integrable chains.  Since the (fused)
transfer matrices are standard objects in the QISM, the relation
(\ref{Qbar}) shows that the $Q$-operator also fits naturally in
the QISM. In contrast to the approach \cite{Nep04}, here it is
presumably essential that the bulk anisotropy parameter $\eta$
have generic values, for which case there exist $U_{q}(sl_{2})$
representations of arbitrary spin.  For the open chain, we have
shown in detail how the $T-Q$ relation, together with some
additional properties, determine the eigenvalues (\ref{result}) of
the transfer matrix and the associated Bethe Ansatz equations
(\ref{BAE-2}).  For a given set of bulk and boundary parameters
satisfying the constraint (\ref{Restr-1}), the eigenvalues
$\L^{(-)}(u)$ and $\L^{(+)}(u)$ {\it together} are expected to
constitute the complete set of eigenvalues of the transfer matrix
$t(u)$ \cite{Nep03}.  Our results complement and also generalize
those obtained in \cite{Nep04} at roots of unity.

It would be interesting to determine the conditions for which the
limit (\ref{Qbar}) exists.  We have seen that, for the open chain
with generic values of $\eta$, (\ref{result}), (\ref{Eign-fuc}) is
a solution if the constraint (\ref{Restr-1}) is satisfied.  This
suggests that, for generic values of $\eta$, the constraint
(\ref{Restr-1}) may be a necessary condition for the existence of
the limit (\ref{Qbar}).  This, in turn, suggests that (again, for
generic values of $\eta$), if the constraint (\ref{Restr-1}) is
not satisfied, then there may not be a $T-Q$ relation -- a most
unusual situation for an integrable model, which merits further
investigation.  It would also be interesting to explicitly
evaluate the $Q$-operator directly from Eq. (\ref{Qbar}).

\vspace{1.6truecm}

\noindent{\it Note added:\/} Some special cases of the constraint
equation (4.17), corresponding to particular values of the
discrete parameters $\epsilon_i$, were noted previously in
\cite{Cao03,Gie04,Gie05}. It was noted in \cite{Gie05} that the
constraint equation corresponds to points where the representation
theory of the two-boundary Temperley-Lieb algebra is
non-semisimple, giving rise to indecomposable representations.

\section*{Acknowledgments}
Financial support from  the Australian Research Council (WLY and YZZ)
and the National Science Foundation under Grant PHY-0244261 (RIN)
is gratefully acknowledged.


\end{document}